\documentclass[preprint,prb,aps,showpacs,showkeys,amsmath]{revtex4-2}
\usepackage{euscript}
\usepackage{graphics}
\usepackage{graphicx}
\usepackage{indentfirst}

\begin{document}

\noindent\fbox{
  \parbox{\textwidth}{
    This article may be downloaded for personal use only. Any other use
    requires prior permission of the author and AIP Publishing. This
    article appeared in ``Physics of Fluids 16, 1811–1813 (2004)",
    and may be found at https://doi.org/10.1063/1.1659817
  }
}

\title{Scalar Probability Density Function mixing models need not comply with the linearity and independence hypothesis.}

\author{Juan Hierro}
\affiliation{LITEC, Consejo Superior de Investigaciones
Cient\'{\i}ficas\\
Mar\'{\i}a de Luna 3, Zaragoza 50018, Spain}
\author{C\'{e}sar Dopazo}
\affiliation{CIEMAT, Centro de Investigaciones Energ\'{e}ticas, Medioambientales y Tecnol\'{o}gicas \\
Avda. Complutense, 22, Madrid 28040, Spain}

\date{\today}

\begin{abstract}
In a mixture of scalar fields undergoing diffusive processes governed by Fick's law, the concentration at each point evolves linearly in the concentrations at all points and independently from the other concentrations, when one considers a finite differences integration of their evolution equations. However, these properties must not necessarily be enforced in Probability Density Function (PDF) models, since they are relaxed when conditional expected values are taken.
\end{abstract}


\maketitle

A mixture of chemical species, without non-linear source terms and subject to Fickian diffusion with constant coefficients, evolves according to (see, for instance, Slattery~\cite{Slattery:1972}).
\begin{equation}
\frac{\partial Y_{\alpha}}{\partial t} + u_i \frac{\partial Y_{\alpha}}
{\partial x_i} = D_{(\alpha)} \frac{\partial^2 Y_{(\alpha)}}
{\partial x_i \partial x_i}
\label{eq:1}
\end{equation}
where $Y_{\alpha}$ denotes the mass fraction of the $\alpha$ chemical species, $u_i$ is the $i$th component of the velocity vector, $x_i$ is the $i$th component of the position vector and $D_{\alpha}$ is the diffusive coefficient of the $\alpha$ chemical species. Einstein's convention of summation over repeated indices, except for those inside parenthesis, is used.

It is straightforward to realize that Eq.(\ref{eq:1}) constitutes a set of linear equations for the mass fractions where the evolution of each mass fraction, $Y_{\alpha}$, is independent of the values of the other ones.

Probability Density Function (PDF) methods have been proved~\cite{Pope:1985,Dopazo:1994} to be a useful tool to model the mixing of scalars, especially for highly non-linear, chemically active species. These models are one-point closures where the mixing term is non-closed. The transport equation of the joint PDF of a set of random variables, $\mathbf{\Gamma}$, associated with the set of physical mass fractions, $\mathbf{Y}$, was first derived by Dopazo and O'Brien~\cite{Dopazo/Brien:1974a} and is given by
\begin{equation}
\frac{\partial P(\mathbf{\Gamma}; t)}{\partial t} = -\frac{\partial}
{\partial \Gamma_{\alpha}} \bigl[ \langle D \nabla^2 Y_{\alpha} | \mathbf{Y} =
\mathbf{\Gamma} \rangle P(\mathbf{\Gamma}; t) \bigr]
\label{eq:2}
\end{equation}
where the scalar and the velocity fields are assumed to be homogeneous and isotropic. In Eq.(\ref{eq:2}), $P(\mathbf{\Gamma};t)$ stands for the joint PDF at time $t$ of the set of all random variables, $\mathbf{\Gamma}$, and $\langle D \nabla^2 Y_{\alpha} | \mathbf{Y} = \mathbf{\Gamma} \rangle$ denotes the expected value of the diffusion of the $\alpha$ chemical species, conditioned on the event that each mass fraction, $Y_{\beta}$, lies between $\Gamma_{\beta}$ and $\Gamma_{\beta} + d\Gamma_{\beta}$. From now on, this conditioned diffusion will be represented by the symbol ${\EuScript D}(\mathbf{\Gamma},t)$.

Due to the linearity and independence properties of the original physical fields, $\mathbf{Y}$, it was proposed by Pope~\cite{Pope:1983,Pope:1985} that any closure model of Eq.(\ref{eq:2}) should satisfy them. It should be taken into account that in this paper, the usual definition of independence and linearity in mathematical textbooks is not considered; instead, the definition given by Pope in the context of PDF formulations is used. This definition states that time variations of each scalar field for a generic stochastic particle should be a linear function of the values of that same scalar field in the set of all the particles of the ensemble \footnote{Of course, the contribution from most of them may be zero and, in fact, should be so in any practical model.}. Moreover, these time variations should be independent of any other scalar field. In a mathematical formulation
\begin{equation}
d\Gamma_{\alpha}^{(n)} = M_{nl} \Gamma_{\alpha}^{(l)} dt \label{eq:301}
\end{equation}
should give the time evolution of the $n$-th sample of the $\alpha$ chemical species, where $M_{nl}$ is a generic matrix.
 
It is worth mentioning that this property is satisfied, for instance, by a numerical integration through finite differences of Eq.(\ref{eq:1}), since partial derivatives are approximated by expressions which are linear in the values of the mass fraction of the derived field at different points. The usual definition of linearity and independence, which is trivially satisfied if Eq. (\ref{eq:2}) is considered as a partial differential equation in $P(\mathbf{\Gamma}; t)$, is not addressed here; the definition given in References 2 and 5 concerns the term ${\EuScript D}(\mathbf{\Gamma},t)$.

However, there are several relatively successful closures of Eq.(\ref{eq:2}) such as the binomial Langevin and the mapping closure~\cite{Dopazo:1994} which do not express ${\EuScript D}(\mathbf{\Gamma},t)$ as a linear function of the modelled fields \footnote{Although mapping closures may be rewritten in that way, see~\cite{Subramanian/Pope:1998}.}. To reconcile both facts, two simple fields will be analytically studied and it will be shown that linearity and independence may be lost in the process of computing conditional expected values. This fact means that models which do not preserve them should not be a priori rejected.

In both examples, only a one-dimensional scalar field with random initial conditions will be considered. This implies that the Laplacian operator is reduced to $\partial^2 / \partial x^2$, where $x$ is the space dimension. It could be argued that this assumption means that one is out of the realm of fluid turbulence. However, if the linearity and indepedent hypothesis does not hold for an exact one-dimensional model problem, it is unlikely that it would be satisfied for three-dimensional velocity driven scalar fields. It could also be mentioned that, in any fully developed turbulent situation, the initial condition is irrelevant. Nevertheless, it is also true that there is always a transient time \footnote{It is usually related to the integral time scale of the flow or eddy turn-over time of the most energetic eddies in the flow.} before the initial condition becomes irrelevant and that, especially in combustion problems, the flame tends to keep  the distribution of scalar fields in a sort of ``unpremixed initial condition'' from the viewpoint of mixing. 

In the first example, one works with a scalar field whose initial distribution of mass fractions is given by
\begin{equation}
Y(x,0) = \frac{1}{2} + \theta_1 \cos(k x + \varphi) + \theta_2 \cos[3 (k x + \varphi)]
\label{eq:3}
\end{equation}
where $\varphi$ is a random-phase factor with a uniform distribution between $0$ and $2 \pi$. The numerical coefficients, $\theta_1 = 5/9$ and $\theta_2 = -1/18$, have been chosen so that this function has a plateau-like shape, with no secondary extremal points, as plotted in Figure 1.

Next, the expected value of the diffusion of this field, conditioned on the field itself, is calculated, as this is the relevant quantity on the right hand side of Eq.(\ref{eq:2}). 

Using the trigonometric relation $\cos(3 \alpha) = 4 \cos^3(\alpha) - 3 \cos(\alpha)$ in Eq.(\ref{eq:3}) in order to obtain the value of $\cos(k x + \varphi)$ as a function of $\Gamma$, the random variable associated with $Y(x,t=0)$, one readily obtains
\begin{equation}
\begin{split}
\Gamma - 1/2 &= \biggl[ \theta_1 - 3 \theta_2 \biggr] \cos(k x + \varphi) \\
&+ 4 \theta_2 \cos^3(k x + \varphi)
\end{split}
\label{eq:5}
\end{equation}
For the sake of simplicity, $\cos(k x + \varphi)$ will be referred to as $z$. Therefore, $z$ is the analytic solution of the following third order polynomial equation
\begin{equation}
z^3 + \frac{\theta_1 - 3 \theta_2}{4 \theta_2} z - \frac{\Gamma - 1/2}{4 \theta_2} = 0
\label{eq:6}
\end{equation}
It is straightforward now to understand why Eq.(\ref{eq:3}) was chosen as an initial condition. It was obtained as a function close to a periodic step function with the restrictions that the resulting calculations could be done analytically (so, one only considers up to third order terms) and that it does not display secondary extremal points between the absolute ones (so, there is only one real solution to Eq.(\ref{eq:6}) for all possible values of $\Gamma$ at any time instant). Step functions are important in real turbulent problems since they represent the typical segregated initial condition of mixing problems and are close to the distribution of burnt/unburnt species in combustion flows. 

It is easy to calculate the conditional expected value of the diffusion of $Y(x,t=0)$ as a function of $z$ which is, in its turn, the analytical solution to Eq.(\ref{eq:6})
\begin{equation}
\bigl \langle D \nabla^2 Y \big | Y = \Gamma \bigr \rangle = -D k^2 [(\theta_1 - 27 \theta_2) z + 36 \theta_2 z^3]
\label{eq:7}
\end{equation}
where $D$ denotes the diffusivity coefficient of the scalar. Eq.(\ref{eq:7}) is plotted in Figure 2 with the arbitrary values: $D=2 \times 10^{-5} m^2 s^{-1}$ and $k=100 m^{-1}$. It is immediately realized that the resulting function is not linear. Any model which represents this non-linear behaviour in a weak sense~\cite{Kloeden/Platen:1992} is equally valid as a closure of the PDF associated with $Y(x,t=0)$. This means that, although a model with a linear dependence upon the field itself is not ruled out, the linearity of closure models is not mandatory.

In the second example, one works with a set of three one-dimensional chemical species whose initial mass fractions are given by
\begin{align}
Y_1(x,0) &= 0.3 + 0.3 \cos(k x + \varphi_1) \label{eq:8} \\
Y_2(x,0) &= 0.2 + 0.2 \cos(l x + \varphi_2) \label{eq:9} \\
Y_3(x,0) &= 1.0 - Y_1(x,0) - Y_2(x,0) \label{eq:10}
\end{align}
where $\varphi_1$ and $\varphi_2$ are independent, random variables with a uniform distribution between $0$ and $2 \pi$. Since the third field, given by Eq.(\ref{eq:10}), is a linear function of the other two fields, it is enough to calculate the evolution of these two fields to fully characterize the system. Both of them are supposed to evolve according to the one-dimensional version of Eq.(\ref{eq:1}) without convection and with the same molecular diffusivities, $D_1 = D_2 = D$.

The conditional expected values of the diffusion of $Y_1$ and $Y_2$ are
\begin{align}
\langle D \nabla^2 Y_1 | Y_1 = \Gamma_1, Y_2 = \Gamma_2 \rangle &= -D k^2 (\Gamma_1 - 0.3) \label{eq:11} \\
\langle D \nabla^2 Y_2 | Y_1 = \Gamma_1, Y_2 = \Gamma_2 \rangle &= -D l^2 (\Gamma_2 - 0.2) \label{eq:12}
\end{align}
which satisfy linearity and independence in sense of Reference 2.

However, both properties should also be satisfied by any linear combination of the original scalar fields~\cite{Subramanian/Pope:1998}. It is straightforward to check that, if one chooses $U_1 = (Y_1 + Y_2) / 2$ and $U_2 = (Y_1 - Y_2) / 2$ with associated random variables $\Xi_1$ and $\Xi_2$, one gets
\begin{align}
\langle D \nabla^2 U_1 | U_1 = \Xi_1, U_2 = \Xi_2 \rangle &= -\frac{D (k^2 + l^2) \Xi_1}{2} - \frac{D (k^2 - l^2) \Xi_2}{2} + 0.15 D k^2 + 0.1 D l^2 \label{eq:13} \\
\langle D \nabla^2 U_2 | U_1 = \Xi_1, U_2 = \Xi_2 \rangle &= -\frac{D (k^2 - l^2) \Xi_1}{2} - \frac{D (k^2 + l^2) \Xi_2}{2} + 0.15 D k^2 - 0.1 D l^2 \label{eq:14}
\end{align}
which satisfy linearity but not independence. The evolution of $\Xi_1$ depends on $\Xi_2$ and that of $\Xi_2$ depends on $\Xi_1$. As in the first example, any linear, independent model which represents this non-independent behaviour, in a weak sense, is valid.

As a final conclusion, linearity and independence properties should not be considered as a must for PDF models, since a multipoint quantity (the Laplacian of the field) is represented in the one-point PDF formulation by a one-point statistics (its expected value conditional upon all the fields). Therefore, valid models may rely either on the multipoint character of the physical quantity (with a linear, independent dependence on several samples of the ensemble), or on the functional dependence of the conditional expected values (with a non-linear, non-independent dependence on the own sample whose evolution is being computed).

\section*{Acknowledgements.}
Juan Hierro would like to thank the Spanish Ministry of Science and Technology for its support through a FPU Fellowship.

\newpage
\section*{References.}
\bibliographystyle{unsrt}

\newpage
\section*{List of figures.}
Fig. 1: Mass fraction of $Y(x,t=0)$ according to Eq.(4).

Fig. 2: Conditional diffusion according to Eq.(7). Time instant: $0.0$ s.

\newpage

\begin{figure}
{

\raggedright 

$Y(x,t=0)$ \hfill

\includegraphics[angle=0,width=85mm]{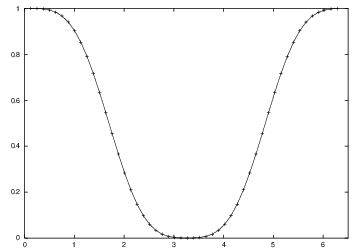}

\hspace*{72mm} $k x+\varphi$ \hfill       

}
\caption{Mass fraction of $Y(x,t=0)$ according to Eq.(4).}
\end{figure}

\vspace*{1cm}

\begin{figure}
{

\raggedright 

${\EuScript D}(\Gamma, t=0)$ \hfill

\includegraphics[angle=0,width=85mm]{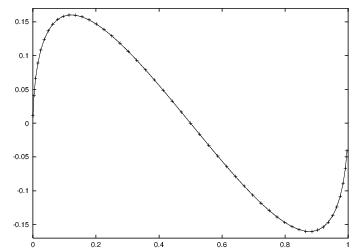}

\hspace*{80mm} $\Gamma$ \hfill

}
\caption{Conditional diffusion according to Eq.(7). Time instant: $0.0$ s.}
\end{figure}

\end{document}